\documentclass[manuscript]{acmart}
\AtBeginDocument{%
  }
\setcopyright{acmlicensed}
\copyrightyear{2025}
\acmYear{2025}

\begin{document}
\title{Theoretical basis for code presentation: A case for cognitive load}

\author{Nyah Speicher}
\email{nspeiche@mail.umw.edu}
\author{Prashant Chandrasekar}
\email{pchandra@umw.edu}
\affiliation{%
  \institution{University of Mary Washington}
  \city{Fredericksburg}
  \state{Virginia}
  \country{USA}
  }

\begin{abstract}
Evidence supports that reducing cognitive load (CL) improves task performance for people of all abilities.
This effect is specifically important for blind-and-low-vision (BLV) individuals because they cannot rely on many common methods of managing CL, which are frequently vision-based techniques.
Current accessible ``solutions'' for BLV developers only sporadically consider CL in their design.
There isn’t a way to know whether CL is being alleviated by them.
Neither do we know if alleviating CL is part of the mechanism behind why these solutions help BLV people.
Using a strong foundation in psychological sciences, we identify aspects of CL that impact performance and learning in programming.
These aspects are then examined when evaluating existing solutions for programming sub-tasks for BLV users.
We propose an initial design ``recommendations'' for presentation of code which, when followed, will reduce cognitive load for BLV developers.
\end{abstract}


\received{11 September 2025}

\maketitle

\section{Introduction}

Writing computer code is a creative expression of problem solving.
Significant development effort for projects in software, web development, machine learning, among others requires an integration of existing libraries or code repositories.
Solutions are naturally built upon one another, re-packaged, extended, re-used, and creatively re-organized to solve different problems.
This creative expression results in scenarios involving unnecessary challenges in programming tasks (such as code comprehension, editing, navigation, among others) for current and aspiring blind-and-low-vision (BLV) developers \cite{Mountapmbeme_Okafor_Ludi_2022}.
We believe that one of the underlying aspects that is core to many of the challenges (as studied extensively in literature) for BLV developers is the impact on cognitive load (CL).

For instance, one of the reasons that code debugging is challenging for BLV developers is because variable ``watch windows'' are inaccessible.
As a result, BLV developers have to use their memory or external processes to keep track of dynamic program behavior and variable states. 
Similarly, BLV developers need to maintain an active memory of multi-level-nested code structures, cursor position, line location, among others, all of which put a strain on their memory, thereby impacting their task performance. 

The main challenge is that the broader developer community is not incentivized to write code in ``a manner'' that might lower the cognitive load for the BLV community.
This makes it continuously challenging for BLV developers to build feature-rich and well integrated solutions.
Even if we could incentivize them, defining such coding ``principles'' or ``rubrics'' that conceptually and practically lower cognitive load is a non-trivial task and needs to based on evaluating the application of psychology-grounded theories.
If defined, these ``theoretically-provable'' considerations can be the foundation for design of future solutions that address accessibility barriers while programming.

\subsection{Research Goal and Research Questions}

The goal of this work is to examine the role that cognitive load plays in the design of various solutions that address accessibility barriers while programming. We,
\begin{itemize}
    \item Identify and discuss the role/importance of CL in design of artifacts/solutions meant to support BLV individuals for \textit{non-programming-related tasks}.
    \item Identify and \textbf{contrast} that against the role of CL in the design of artifacts or solutions meant to support BLV individuals for \textit{programming-related} tasks.
    \item Thereby, identify gaps in current research on designing accessible solutions keeping CL as a criteria.
    \item Propose rules for code structure which are composed with consideration for CL and that have a significant impact on performance of downstream programming tasks for BLV individuals.
\end{itemize}

To achieve the goals, we formulated the following research questions:
\begin{itemize}
    \item \textbf{RQ1:} How does cognitive load (CL) factor into the design of solutions meant to support BLV individuals?
    \item \textbf{RQ2:} What existing solutions/approaches meant to support programming tasks (in-)directly consider CL in their designs?
    \item \textbf{RQ3:} In what ways do solutions/designs for programming tasks which consider CL improve upon those that don't consider CL?
\end{itemize}

We first provide a background on aspects of cognitive load theory and the emphasis that psychologists put on its impact on task performance (Sections \ref{sec:cog_load_working_memory} and \ref{sec:types_cog_load}).
We provide our review of the impact of cognitive load in the design of solutions meant to support BLV individuals for non-programming tasks (Section \ref{sec:non_prog_cog}).
We conduct an exhaustive evaluation of all existing solutions meant to support BLV developers in their programming tasks (Section \ref{sec:prog_cog}). 
Specifically, we inspect the rationale behind the solution designs and check for basis in cognitive load theory.
Using our expertise in Psychology, CS-Education, and Accessibility in Computing, we synthesize our findings and propose the creation of a ``virtual code view'' that is a projection of any code, refactored, to reduce the burden of cognitive load (Section \ref{sec:virtual_code}).

\section{Cognitive Load and Working Memory}
\label{sec:cog_load_working_memory}
Cognitive load (CL) encompasses the demand placed on an individual's mental resources by a particular task.
It is a term to describe how those resources are stretched in order to achieve a result \cite{Sweller_1988, Debue_Van_De_Leemput_2014}.
Understanding the role of CL is important because it is associated with poorer performance on measures of learning and task execution \cite{Sweller_1988, Krigolson_Heinekey_Kent_Handy_2012}.

In order to understand CL, one must understand working memory, because the two concepts are inextricably linked.
Working memory is a part of the short-term memory system.
It serves as a sort of ``workbench'' for information that is immediately relevant to the task at hand.
While information is in working memory, it can be manipulated and used to make sense of processes such as learning, comprehension, or reasoning \cite{Cowan_2010, Adam_Vogel_Awh_2017}.
Information generally moves in and out of working memory quickly unless it is consciously maintained by the individual.
Even with conscious maintenance, information in working memory is fleeting because the brain is driven by novel stimuli \cite{Caulfield_Zhu_McAuley_Servatius_2016, Ranganath_Rainer_2003}.
Human attention is easily stolen away from one focus by a new object of interest in the environment.
In this case, the information previously contained in working memory is quickly abandoned to make room for new information.

\textbf{Importantly, working memory only has a limited capacity.}
Recent studies place this between just three and five ``chunks'' of information at a time \cite{Cowan_2010, Adam_Vogel_Awh_2017}.
For this reason, it is a precious resource which the human brain has to manage deliberately.
When participating in a goal-directed activity, there is a careful balance between allocating working memory resources to productive processes and being cautious not to overload the working memory and cause all of the individual's mental resources to be depleted.
It has been demonstrated that overloading the brain's cognitive resources, aptly named \textbf{cognitive overload}, is associated with \textbf{poorer performance} and \textbf{learning outcomes} \cite{Sweller_1988, Krigolson_Heinekey_Kent_Handy_2012}.
Alternatively, there is evidence that maintaining a balance of the working memory and other cognitive resources contributes to improved performance \cite{Debue_Van_De_Leemput_2014}.

A great example of how high CL impedes to processes of learning and task performance comes from a study in which the authors looked at the impact of different CL levels on the ability to evaluate and adjust to error feedback \cite{Krigolson_Heinekey_Kent_Handy_2012}.
The procedure involved participants performing a time estimation task, and then receiving feedback about their performance on the task.
Using that feedback, they would then attempt to improve their accuracy on the next trial.
The authors found that participants were able to make more improvement between trials in the low CL condition compared to the low CL condition.
This indicates that they were less effective at using error feedback to make adjustments in the high CL condition.
In other words, they weren't able to learn as well from feedback during past trials.
Although it only only looks at one type of learning, this study demonstrates how CL plays a key role in learning, memory, and performance.

\section{Types of Cognitive Load}
\label{sec:types_cog_load}
Cognitive load is made up of not one, but three interdependent components.
Each of these types of CL contributes to the overall demand a person experiences, but they come from unique sources.
It is important to understand the mechanisms behind these CL elements because they play different roles in the greater impact of CL.
For example, some elements of CL can be modified by external means, while others are fixed entities.
Knowing what part each of them plays in the context of learning and task performance enables them to be addressed effectively.

\subsection{Intrinsic Cognitive Load}

Intrinsic cognitive load (ICL) is the first of the three CL components.
It comprises the mental effort expended by a person due to the difficulty or complexity of the task at hand \cite{Bannert_2002, Issever_Catalbas_Duran_2023}.
ICL cannot be modified by outside influences because it is inherent to the material.
However, it is moderated by the expertise of the individual.
A higher level of expertise comes with more experience and better strategies for processing information which don't rely as heavily on the working memory system.
I.e., experts are better at coping with complex material efficiently, so they don't experience the same volume of ICL.

\subsection{Extraneous Cognitive Load}

Extraneous cognitive load (ECL) makes up the second portion of CL.
It describes the excess cognitive resources devoted to processing information indirectly related to the task goal (e.g., distracting stimuli, irrelevant instructions, confusing wording) \cite{Bannert_2002, Issever_Catalbas_Duran_2023}.
ECL can be problematic because these distracting stimuli, which do not contribute to the task goals, draw attention and resources away from productive learning processes.
\textbf{This is why, in both learning and task performance, minimizing ECL is associated with better outcomes} \cite{Bannert_2002}.
Reducing ECL redirects mental resources towards productive, goal-directed processes.
ECL is important in the realm of design because it is the \textbf{element of CL which can most easily be externally modified}.
This is particularly true when ICL is high because these components are additive.
High ICL leaves few resources to spare, so designs which minimize ECL facilitate the efficient use of cognitive resources \cite{Bannert_2002, Issever_Catalbas_Duran_2023}.

\subsection{Germane Cognitive Load}
Germane cognitive load (GCL) is the final element of CL.
While it is often the most difficult to conceptualize, it is also the most important.
It is easiest to think about GCL as the mental effort directed towards goal-related information processing.
This can involve making connections between old and new knowledge as well as applying learning strategies to understand information \cite{Bannert_2002, Issever_Catalbas_Duran_2023}.
In a way, GCL can be described as ``good'' CL.
Although it is still ideal to keep CL low, if a large portion of the general load is made up of GCL, then those cognitive resources are being utilized productively.
High GCL can be an indication that the individual is using their mental effort towards learning strategies, attention, and memory, rather than distracting stimuli as in ECL \cite{Debue_Van_De_Leemput_2014}.
This type of load can be externally influenced to a limited extent by teaching learning strategies and schema construction, but the effect is not as substantial as that of focusing on ECL \cite{Bannert_2002}.

\section{Cognitive Offloading}
Cognitive offloading is defined as ``the use of physical action to alter the information processing requirements of a task so as to reduce cognitive demand'' \cite{Risko_Gilbert_2016}.
It is a mechanism which enables people to work around the limitations of mental resources like working memory and perception.
The idea of offloading is that a physical action takes on some of the demand of the task so there is less demand placed on the cognitive resources of the individual.
As discussed, working memory capacity is limited, and the same is true about other cognitive resources
It is not always feasible to change information presentation or clarify instructions on a large scale in order to minimize CL.
This is why developing strategies on the individual level to cope with excessive CL is so beneficial.
Offloading provides a method for people to manage CL levels in specific contexts.

There are a variety of methods for cognitive offloading, but the most important category to understand in this case is called ``into-the-world''.
This term refers to the type of tool the processing demand is being offloaded onto in the method.
``Onto-the body'' is second method in which some cognitive requirements of the task are absorbed by a physical action of the body \cite{Risko_Gilbert_2016}.
For example, someone counting on their fingers or tilting their head to view a crooked image would fall under the category of ``onto-the-body''.
However, many of the offloading strategies which are most easily recognized fall under ``into-the-world''.
Offloading into-the-world involves the use of a tool or device to create an external representation of information.
Rather than an individual having to store the information in memory, it is stored in a physical, separate form which utilizes no (or fewer) cognitive resources \cite{Risko_Gilbert_2016}.
Additionally, offloading into-the-world has the benefit of being helpful for both retrospective memory (remembering events, people, experiences, or information from the past) as well as prospective memory (remembering to complete a future planned action).
Both of these types of memory are essential for task completion.
Retrospective memory allows us to remember all of the relevant experiences and information that inform how we accomplish a goal.
However, prospective memory is just as important for prioritizing and managing workflow.
Offloading provides techniques to better manage both retrospective and prospective memory requirements within a task.

Cognitive offloading is beneficial for anyone, regardless of background or abilities.
It has been demonstrated to reduce cognitive load and improve performance on a variety of tasks \cite{Risko_Gilbert_2016}.
However, this strategy is not as accessible to blind and low vision (BLV) individuals as it is to sighted individuals.
\textbf{The primary reason is that the most commonly used tools for cognitive offloading are vision-based in design.}
They include devices like paper calendars and agendas, written lists, and google calendar \cite{Baptista_Afonso_Silva_2023}.
Although some of these tools can still be used by BLV individuals, many of their most helpful features are inaccessible to someone with impaired vision.
For this reason, BLV individuals face unique challenges with regard to managing CL compared to sighted people.
This is partially because BLV individuals have fewer offloading options available to manage their cognitive resources productively.

\section{Cognitive Load in Design of Accessible Solutions} 
Interestingly, it is specifically the process of making working memory space available which appears to cause performance increases \cite{Li_Frischkorn_Dames_Oberauer_2025}.
The act of offloading information from working memory to external memory aids is the most common form of this operation in everyday life.
There has long been uncertainty about the mechanisms behind removing information from working memory leading to performance benefits.
Some scholars believed it was primarily due to reduced proactive interference (the action of prior information in memory inhibiting the ability to learn new information).
However, other researchers thought it was more likely a result of freeing up space in working memory which had previously been occupied by those items.
Most recent findings suggest that working memory capacity is the more probable of these two explanations, although both doubtlessly play a role in the complicated cognitive interaction \cite{Li_Frischkorn_Dames_Oberauer_2025}.

It is for this reason that researchers should consider cognitive load and cognitive offloading in the design of accessibility tools for BLV individuals.
In some areas of accessibility research, this is already common practice. However, it is still an exception in many accessible programming tools.
We will explore the state of existing accessible programming tools, a majority of which do not examine cognitive load in their design.
We will also discuss how these tools can improve accessibility further by considering cognitive load and providing tools for cognitive offloading within their models.

\subsection{Non-Programming Tasks}
\label{sec:non_prog_cog}
It would be a stretch to say that designing with CL in mind is the norm in BLV accessibility research, but it is true that many more tools for BLV accessibility have taken it into account in recent years.
Especially for accessibility solutions outside of programming, there are a number of tools which factor CL into their design.
A great, and recent, example of this is the Aerial Guide Dog, a tool to help BLV individuals with navigation indoors \cite{Zhang_Pan_Song_Zhang_Li_Ding_2024}.
Early on in the design process of the navigation tool, the authors address that many navigation solutions provide an overwhelming amount of information to the user.
The many cues and sensory signals can require a huge amount of mental effort to decode, and are coupled with a pretty steep learning curve.
In some cases, this can look like having to learn the meaning of a dozen different auditory cues.
This far exceeds the capacity of an average person's working memory, so it's very difficult to learn quickly when an individual has not yet established any strategies or schemas to store the information effectively outside the working memory.
The Aerial Guide Dog design instead takes advantage of instinctive tactile senses by physically guiding the user, much like a real guide dog.
The idea is that this requires far less conscious mental effort from the user and frees up resources for other processes.

Similarly, a 2022 study evaluated BLV users interacting with a computer using different methods from a CL perspective \cite{Modanwal_Rai_Jaiswal_Singh_Sarawadekar_2022}
One group in the study used braille to interact with the computer, and the other used dactylology, a method utilizing gestures for computer interaction.
There were also three conditions for CL (low, medium, high) mediated by task complexity (ICL) which allowed the researchers to compare performance on the task at different levels of load.
Their goal was to identify if the pattern of performance between the interaction methods differed at different levels of CL.
Performance was measured based on response time and false responses.

These are just a couple of the accessibility tools for BLV individuals in the last few years which have considered CL in their design, but they highlight the point that CL is an important factor.
Particularly for BLV people, designing tools which minimize CL is critical to optimizing performance for the people using those tools.

\subsection{Programming Tasks}
\label{sec:prog_cog}
In order to evaluate the current level of consideration for CL in existing programming accessibility solutions, we assessed a collection of recent tools developed in that area.
The tools evaluated for CL consideration come from a 2022 literature review \cite{Mountapmbeme_Okafor_Ludi_2022}.
In the review, the authors give an overview of some of these existing accessibility tools and discuss the coding challenge addressed by each of them.
We evaluated the same list of tools for our purposes.
The results of the review are found in Table \ref{tab:accessible_programming_tools}.
The programming challenge addressed by each tool, as identified in the original paper, is included.
The new addition to the table is the final column which recognizes the extent to which CL was considered in the design of the solution.
Tools listed as `No' did not include any discussion of CL in the design rationale or any evaluation of CL when measuring participant performance with and without the tool. 
Those listed as `Yes' explicitly mentioned CL-related concepts such as the difficulty of distinguishing many different commands or wanting to avoid overwhelming users with excessive information. 
Finally, the `Maybe' designation is given to those tools which either discussed CL-adjacent concepts, but did not incorporate them into the tool design, or included elements in the tool design which were intended to reduce CL, but did not mention the motivation behind the choice.

The design of accessible programming tools for BLV individuals seems to be a little behind non-programming design in regards to consideration for CL.
Some of these accessibility solutions address CL in their design process, but it is not common.
This is important because evidence supports that CL plays a role in learning and task performance.
Also because it represents a gap between programming and non-programming accessibility tool design.
This known link between CL, learning, and task performance suggests that it would be beneficial to the area of BLV accessibility research to consider CL in studies.
Quantifying the influence of CL on the effectiveness of accessibility tools can deepen the understanding researchers have and open doors to improved designs.

\begin{table*}
    \caption{Cognitive Load in Design of Accessible Programming Tools}
    \label{tab:accessible_programming_tools}
    \begin{tabular}{|l|l|p{6cm}|l|}
        \toprule
        Tool Name & Year & Programming Challenge & CL?\\
        \midrule
        ACONT \cite{Hutchinson_Metatla_2018} & 2018 & Code Navigation, Code Skimming & No\\ \hline
        APL \cite{Sánchez_Aguayo_2006, Sánchez_Aguayo_2005} & 2005 & Inaccessibility of programming languages & Yes\\ \hline
        AudioHighlight \cite{Armaly_Rodeghero_McMillan_2018} & 2018 & Code Navigation, Code Skimming & Maybe\\ \hline
        Aural Tree Navigator \cite{Smith_Cook_Francioni_Hossain_Anwar_Rahman_2003} & 2003 & Code Navigation, Code Skimming & No\\ \hline
        Blockly (Accessible Version; Caraco et al.) \cite{Caraco_Deibel_Ma_Milne_2019} & 2019 & Inaccessibility of BBLs & No\\ \hline
        Blockly (Accessible Version; Ludi et al.) \cite{Ludi_Spencer_2017} & 2015 & Inaccessibility of BBLs & No\\ \hline
        Blocks4All \cite{Milne_Ladner_2019, Milne_Ladner_2018, Ong_Amoah_Garrett-Engele_Page_McCarthy_Milne_2019} & 2018 & Inaccessibility of BBLs & No\\ \hline
        Bonk \cite{Kane_Koushik_Muehlbradt_2018} & 2018 & Inaccessibility of programming through media & No\\ \hline
        COBRIX \cite{Ahn_Sung_Lee_I_2017} & 2017 & Inaccessibility of programming tools & No\\ \hline
        Code Mirror Block \cite{Schanzer_Bahram_Krishnamurthi_2019} & 2019 & Code Navigation, Code Editing & Maybe\\ \hline
        CodeBox64 \cite{Wang_Wagner_2019} & 2017 & Inaccessibility of BBLs & No\\ \hline
        CodeRhythm \cite{Rong_Chan_Chen_Zhu_2020_CodeRhythm, Rong_Chan_Chen_Zhu_2020_Designing} & 2020 & Challenges of tangible programming environments & No\\ \hline
        CodeTalk \cite{Potluri_Vaithilingam_Iyengar_Vidya_Swaminathan_Srinivasa_2018} & 2018 & Code Navigation, Code Comprehension, Code Editing, Code Debugging & No\\ \hline
        Donnie \cite{Marques_Einloft_Bergamin_Marek_Maidana_Campos_Manssour_Amory_2017} & 2017 & Inaccessibility of programming languages and robotics toolkit & No\\ \hline
        GUIDL \cite{Konecki_2012} & 2012 & Inaccessibility of GUI-based applications & No\\ \hline
        JavaSpeak \cite{Smith_Francioni_Matzek_2000} & 2000 & Code Navigation, Code Skimming & No\\ \hline
        JBrick \cite{Ludi_Abadi_Fujiki_Sankaran_Herzberg_2010} & 2010 & Inaccessibility of robotics programming environments & Maybe\\ \hline
        Music Blocks \cite{Sabuncuoglu_2020} & 2020 & Inaccessibility of BBLs & Maybe\\ \hline
        Noodle \cite{Lewis_2014} & 2014 & Inaccessibility of visual programming systems & Yes\\ \hline
        P-CUBE \cite{Kakehashi_Motoyoshi_Koyanagi_Ohshima_Kawakami_2013} & 2013 & Inaccessibility of BBLs & No\\ \hline
        Phogo \cite{Molins-Ruano_Gonzalez-Sacristan_Garcia-Saura_2018} & 2017 & Inaccessibility of programming languages & No\\ \hline
        Pseudospatial Blocks \cite{Koushik_Lewis_2016} & 2016 & Inaccessibility of BBLs & No\\ \hline
        Robbie \cite{Howard_Hyuk_Park_Remy_2012, Remy_2013} & 2012 & Inaccessibility of visual feedback produced by robots & Maybe\\ \hline
        Scripting Language \cite{Sánchez_Aguayo_2005, Siegfried_Diakoniarakis_Franqueiro_Jain_2005_Extending, Franqueiro_Siegfried_2006, Siegfried_Diakoniarakis_Obianyo-Agu_2005_Teaching, Siegfried_2006, Siegfried_2002} & 2002 & Inaccessibility of GUI-based applications & Maybe\\ \hline
        SODBeans \cite{Stefik_Haywood_Mansoor_Dunda_Garcia_2009, Stefik_Hundhausen_Smith_2011, Stefik_2008} & 2008 & Code Debugging & No\\ \hline
        Sparsha \cite{Shetty_2020} & 2020 & Editing visual layout templates & No\\ \hline
        Story Blocks \cite{Koushik_Kane_2017, Koushik_Guinness_Kane_2019} & 2017 & Inaccessibility of BBLs & No\\ \hline
        StructJumper \cite{Baker_Milne_Ladner_2015} & 2015 & Code Navigation, Code Skimming & Maybe\\ \hline
        Tactile Code Skimmer \cite{Falase_Siu_Follmer_2019} & 2019 & Code Navigation, Code Skimming & Yes\\ \hline
        Tangible Programming Tool Prototype \cite{Utreras_Pontelli_2020} & 2020 & Inaccessibility of BBLs & No\\ \hline
        TIP-Toy \cite{Barbareschi_Costanza_Holloway_2020} & 2020 & Inaccessibility of BBLs & Maybe\\ \hline
        Torino \cite{Villar_Morrison_Cletheroe_Regan_Thieme_Saul_2019, Thieme_Morrison_Villar_Grayson_Lindley_2017, Morrison_Villar_Thieme_Ashktorab_et_al._2020} & 2017 & Inaccessibility of BBLs & Maybe\\ \hline
        Wicked Audio Debugger \cite{Stefik_Alexander_Patterson_Brown_2007} & 2007 & Code Debugging & Maybe\\
        \bottomrule
    \end{tabular}
\end{table*}

\section{Opportunity: Creating a ``Virtual Code View''}
\label{sec:virtual_code}

In SQL, a \textit{view} is a virtual table based on the result-set of an SQL statement. 
It presents data in rows and columns, similar to a real table, but it does not store the data itself. Instead, a view's data is dynamically generated by executing the underlying SQL query whenever the view is accessed.

In the same vein, we propose the need of a ``virtual code view'' for BLV developers.
This ``view'' presents code in a manner that is a ``relatively lighter'' load for BLV developers when performing downstream programming tasks.
The initial set of criteria for the refactoring and creating this ``view'' are:
\begin{itemize}
    \item \textbf{Flattening Code}: Nested methods, conditionals, or control statements are creative choices that impact BLV developers for code navigation-based tasks. We would refactor the code to ``flatten it'' such that navigating the code requires less of jumping around and jumping within.

    \item \textbf{Minimize variable or function call stack size}: As we pointed out, it is ideal for task performance if the working memory has to store (or recall) 3-5 items of information. We would refactor the code such that the combination of the size of the call stack, responsible for keeping track of functions call, and scope of variables, is minimized to lowest possible.

    \item \textbf{Character limit for each line}: There are a small subset of BLV developers who rely on external devices, such as Braille displays, for code comprehension. There is no single ``standard'' size for Braille displays. One can find displays with up to 20 braille cells and up to 40 braille cells. Having lines of code span more than 40 characters causes an inconvenience because it requires storing (or recalling) prior information. We would refactor the code to have a upper bound of 40 characters per line. 
\end{itemize}

These would be the ``version 1.0'' of guidelines for ``Code Accessibility'' that LLMs could use to generate ``code views'' with lighter cognitive load.
With a ``virtual view'' based on cognitive load theory, we are preparing to run a user study to evaluate task performance across different sub-tasks.

\section{Conclusion}

This effort is the first step and we plan to extend and/or modify virtual code view based on other psychological, and environmental factors. 
As always, any changes to our conceptual design will continuously be tested with BLV developers.
Given the fundamental importance of the role of cognitive load in the performance of tasks, it is surprising that we discovered its sporadic use as a measure for evaluating or a basis for designing solutions to support programming for BLV individuals.
Through our detailed review of existing solutions, we highlight shortcomings and, thereby, a gaping opportunity to present BLV developers the tools that would make all of programming significantly more accessible.

\bibliographystyle{ACM-Reference-Format}
\bibliography{sample-base}

\end{document}